# Title: Topological surface states in γ-PtBi$_2$ evidenced by scanning tunneling microscopy

**Authors:** Yunkai Guo[a,b], Jingming Yan[a,b], Wen-Han Dong[a,b], Yongkai Li[c], Yucong Peng[a,b], Xuetao Di[c,d,e], Caizhen Li[c,d,e], Zhiwei Wang[c,d,e*], Yong Xu[a,b], Peizhe Tang[f], Yugui Yao[c,d,e], Wenhui Duan[a,b], Qi-Kun Xue[a,b,g,h,i *], Wei Li[a,b,i*]

[a]*State Key Laboratory of Low-Dimensional Quantum Physics, Department of Physics, Tsinghua University, Beijing 100084, China*

[b]*Frontier Science Center for Quantum Information, Beijing 100084, China*

[c]*Centre for Quantum Physics, Key Laboratory of Advanced Optoelectronic Quantum Architecture and Measurement (MOE), School of Physics, Beijing Institute of Technology, Beijing 100081, China*

[d]*Beijing Key Lab of Nanophotonics and Ultrafine Optoelectronic Systems, Beijing Institute of Technology, Beijing 100081, China*

[e]*International Center for Quantum Materials，Beijing Institute of Technology, Zhuhai 519000, China*

[f]*School of Materials Science and Engineering, Beihang University, Beijing 100191, China*

[g]*Beijing Academy of Quantum Information Sciences, Beijing 100193, China*

[h]*Southern University of Science and Technology, Shenzhen 518055, China*

[i]*Hefei National Laboratory, Hefei 230088, China*

*To whom correspondence should be addressed: zhiweiwang@bit.edu.cn; qkxue@mail.tsinghua.edu.cn; weili83@tsinghua.edu.cn;

Trigonal PtBi₂ (γ-PtBi₂) is a newly discovered polymorphic material with novel topological and superconducting properties [1, 2]. Due to its non-centrosymmetric structure, γ-PtBi₂ hosts triply degenerate points and Weyl fermions [3-6]. On the other hand, bulk superconductivity in γ-PtBi₂ has been observed with a critical temperature $T_c$ up to 0.6 K [7]. Recent studies also uncover its surface superconductivity with $T_c$ of 10 K [6, 8]. Therefore, γ-PtBi₂, possessing non-trivial electronic topology and superconductivity, is a potential candidate for realizing the Fu-Kane topological superconductivity [6, 9, 10], which is timely needed to be further studied by low-temperature scanning tunneling microscopy (STM). However, as the first step, STM evidences of the topological surface states in γ-PtB₂ are still lacking.

In this letter, we report multiple STM evidences for the topological surface states in γ-PtBi₂. More importantly, we demonstrate that the topological surface states are precisely located over a narrow energy range near the Fermi level, which is essential for realizing the Fu-Kane topological superconductivity.

γ-PtBi₂ is a van der Waals material with the space group P31m, which displays Bi-Pt-Bi stacking as shown in Fig. 1a. Due to the lack of inversion symmetry, γ-PtBi₂ possesses different Bi layers as the top and bottom Bi, thus having two types of cleaved Bi-surfaces. In our experiments, the atomically resolved STM topography shows a honeycomb structure with the in-plane lattice constant of 0.66 nm (Fig. S1a), corresponding to the top Bi surface (see Fig. 1a) [5, 10]. We can barely find point defects on the surface, indicating the high quality of the sample. Figure 1b shows an area of γ-PtBi₂ sample, where several step-edges and screw dislocations can be observed. Those edges and dislocations can act as scattering sources for quasiparticle interference (QPI), illustrated by the sky blue and yellow ripples in Fig. 1b. STM can directly measure the QPI and then investigate the topological surface states. Due to the low number of point defects, we primarily focus on the scattering caused by the step-edge and the screw dislocations.

Figure 1d and 1e shows d$I$/d$V$ mappings taken in the vicinity of a screw dislocation and step edge in Fig. 1b. Elegant ripple patterns are clearly visualized at different energies, indicating possible surface states interference. The strong QPI signal is a characteristic of the two-dimensional surface states [11], while the three-dimensional bands have strong dispersion in the $k_z$ direction, causing the scattering vector to be averaged out, which prevents the generation of strong signals [12]. As seen in the Figure 1d, the screw dislocation can select two specific directions marked by the white arrows with a 120-degree angle in real space. The signal along the dashed arrow direction is weaker, so we mainly focus on the stronger direction. Full view of the screw dislocation and its QPI are shown in Fig. S1b and c. In contrast, the QPI shown in Fig 1e is parallel to the step-edge. We perform fast Fourier transform (FFT) to those mappings and obtain the corresponding scattering wave vectors in scattering space. At energy of 0 meV, a scattering wave vector $q_1$ ($q_2$), the sharp features in the yellow (sky blue) circles in Fig. 1f, is extracted from the FFT image. The signals in yellow circles do not appear as points because they reflect the shape of the Fermi arcs. The $q_2$ obtained from the FFT corresponds well to the periodicity of the QPI in real space (Fig. S2), while the features inside the gray dashed box arise from surface topography imperfections and are not associated with the observed QPI signals.

To investigate the origin of the scattering wave vector $q_1$ and $q_2$, we calculate the band structure of the surface state of γ-PtBi₂ in $k$-space with the termination of the top Bi surface in the framework of density functional theory (DFT) and plot the constant energy contours (CECs) at the Fermi level ($E_F$) with spin textures (Fig. 1c). In consistent with previous studies [5, 6], we confirm the topological nature of γ-PtBi₂ as the Weyl semimetal and observe the Fermi arc surface states. Via comparing the calculated CECs with the measured $q_1$ and $q_2$, we find that they mainly originate from the scattering between the Fermi arcs along the Γ-M and Γ-K direction (see the yellow and sky-blue arrows in Fig. 1c). Moreover, the scattering wave vector $q_1$ obtained from our measurements corresponds well with the high-resolution Fermi surface map from angle-resolved photoemission spectroscopy [6], indicating that it indeed originates from the scattering between the Fermi arcs (see Fig. S3). Intriguingly, only the scattering vector $q_1$ appears in our mappings along the Γ-M direction, and the wave vector related to the back-scattering process (denoted by the gray arrow in Fig. 1c) is absent. This can be understood if we consider the spin-texture of the related CECs. Based on our DFT calculation, due to the time reversal symmetry, two states with opposite momenta [$\boldsymbol{\psi}(k)$ and $\boldsymbol{\psi}(-k)$] should host completely opposite spin textures. The direct back-scattering process is forbidden, since the non-magnetic defects cannot induce such kind of spin-flip scattering processes. As for $q_1$, the parallel component of the spin allows scattering to occur. Therefore, the topological nature of the Fermi arcs, reflected in their spin characteristics, is clearly verified by our QPI results. Similar phenomena have been observed in the QPI of topological insulators [11].

To further confirm the contribution of the topological surface states to the QPI signals, we take the energy-dependent line profiles crossing the $q_1$ (along $q_{Bragg}$) and $q_2$ in Fig. 1f. The intensity of each line profile at specific energy is normalized by the corresponding intensity of the Bragg peak, and the normalized results are shown in the upper panel of Fig. 1g and 1h. The signals with almost no dispersion near $k = 0$ correspond to the morphology of the scattering source. As shown in Fig. 1g, the signals at the outermost sides, exhibiting almost non-dispersive behavior, correspond to the Bragg peaks, while the signals indicated by the yellow arrows on the inner sides, exhibiting weak dispersion, correspond to the $q_1$. Surprisingly, the normalized intensities of $q_1$ and $q_2$ are strongly enhanced near the $E_F$. The same phenomenon is also observed for $q_2$ in Fig. 1h. For a more detailed analysis, we take linecuts of $q_1$ and $q_2$ along the direction of energy dispersion (see the yellow and sky-blue arrow in upper panel of Fig. 1g and 1h) to examine the variation in its intensity with energy. As shown in Fig. 1i, the intensity of $q_1$ and $q_2$ exhibits a peak-like behavior within the same energy range near the $E_F$. The peak features are more pronounced on the left side of the Fermi energy, which is consistent with the ARPES measurements of the Fermi arcs, as shown in Fig. S4. Additionally, our calculations support that the Fermi arc states are prominent in a narrow energy range around the $E_F$ (see Fig. S5). These together indicate that the topological surface states dominate the QPI signal in the blue-shaded region around the $E_F$, while the bulk electrons participate the scattering far from the $E_F$ and give rise to the decreased intensity of $q_1$ ($q_2$). An increase in quasiparticle lifetime can also lead to signal enhancement [13]. The quasiparticle lifetime is related to the degree of signal broadening, specifically the full width at half maximum (FWHM), as shown in Fig. S6. As the signal moves from far from the Fermi level to near the Fermi level, the FWHM remains constant or even increases. This rules out the possibility that the signal enhancement is caused by an increase in quasiparticle lifetime.

Up to now, the main scattering wave vectors between the surface states of the Fermi arcs of γ-PtBi$_2$ have been observed in our experiments. Moreover, the suppression of back-scattering and the enhancement of the scattering of $q_1$ and $q_2$ indicate the existence of topologically non-trivial Fermi arc surface states with exotic spin textures on γ-PtBi$_2$.

The spatial decay of the local density of states (LDOSs) oscillations can also exhibit the topological nature of the surface states [14]. We extract the LDOSs of two types of QPI at different energies (see Fig. 2a and Fig. 2b). The LDOSs are expressed as a function of distance from the scattering sources and thus can be fitted by the formula LDOS $\propto \cos(kx + \varphi)x^r$, where $k$ is scattering wave vector in Fig. 1g or Fig. 1h, $x$ is the distance from the scattering source, $\varphi$ is the phase, and r is the spatial decay index that indicates the rate of decay [14]. The electron scatterings of bulk states and surface states have different decay indices. In γ-PtBi$_2$, the strong coupling between surface states and bulk states make it relatively difficult to obtain the exact value of decay index. However, the decay index can still indicate whether the contribution of scattering from the surface states or not. If the scattering mainly arises from the topological surface states, the LDOSs should decay more slowly. As shown in Fig. 2c, the trend of the decay indices as a function of energies has a similar peak-like feature to those observed in the QPI intensities, demonstrating that both the QPI patterns decay more slowly near the $E_F$ than those away from $E_F$. And the energy range here where the peak appears is consistent with those of the intensity enhancement of QPI (Fig. 1i). From another perspective, it further corroborates that the topological surface states indeed precisely locate near the $E_F$.

In summary, using STM, we observe clear QPI patterns in γ-PtBi$_2$. Such a strong signal can only originate from the surface states. Within a similar narrow energy range near the $E_F$, the intensities of QPI are enhanced, and slowly decay in real space, demonstrating that the surface states indeed exist and locate precisely near the $E_F$. Furthermore, the lack of back-scattering makes the observed topologically non-trivial QPI more remarkable, compared with the conventional surface states observed on the metal samples [15]. Recently, the enhanced surface superconductivity has been observed in γ-PtBi$_2$ [6, 8, 10], and the maximal gap size can reach up to 20 meV. In our sample, we only observe a dip-like feature in the density of states near the $E_F$ (see Fig. S7). It is not symmetric with respect to the $E_F$ and not sensitive to magnetic field. Thus, we speculate that it may not be a superconducting gap. Our transport measurements show hints of bulk superconductivity around 0.65 K, but the transition is relatively broad, and the resistance does not drop to zero (Fig. S8c). This behavior is consistent with what has been reported in previous literature [7]. Therefore, we have not observed decent superconducting gap in tunneling spectra. In addition, the previously reported enhancement of surface superconductivity [10] is not observed in our samples. Our electrical resistivity and X-ray diffraction measurements indicate that the sample possesses relatively good quality (Fig. S8a, b). Therefore, the conditions required for the emergence of bulk superconductivity as well as surface superconductivity enhancement need to be further explored.

## Conflict of Interest

The authors declare no competing interests.

## Acknowledgements

The research was supported by the National Natural Science Foundation of China (Grants Nos. 92365201, 12234011, 12374053, 12334003, 52388201), the Ministry of Science and Technology of the People's Republic of China (Grant No. 2022YFA1403100, 2023YFA1406400), the Innovation Program for Quantum Science and Technology (2021ZD0302402), the National Key Research and Development Program of China (Grant Nos. 2022YFA1403400, 2020YFA0308800), the Beijing National Laboratory for Condensed Matter Physics (Grant No. 2023BNLCMPKF007) and the Beijing Natural Science Foundation (Grant No. Z210006). Yong Xu and Wenhui Duan were supported by Innovation Program for Quantum Science and Technology (Grant No. 2023ZD0300500), the National Science Fund for Distinguished Young

Scholars (Grant No. 12025405), the Beijing Advanced Innovation Center for Future Chip (ICFC), and the Beijing Advanced Innovation Center for Materials Genome Engineering.

**Author contributions**

Wei Li and Qi-Kun Xue conceived and supervised the research project. Yunkai Guo, Jingming Yan and Yucong Peng performed the STM experiments. Yongkai Li, Zhiwei Wang and Yugui Yao grew the samples. Xuetao Di, Caizhen Li, Zhiwei Wang and Yugui Yao performed electrical resistivity and X-ray diffraction measurements. Wei Li, Yunkai Guo and Jingming Yan analyzed the data. Wen-Han Dong, Peizhe Tang, Yong Xu and Wenhui Duan carried out the DFT calculations. Wei Li and Yunkai Guo wrote the manuscript with input from all other authors.

**Data availability**

The data that support the findings of this study are included in this article and its supplementary information file and are available from the corresponding author upon reasonable request.

**Appendix A. Supplementary materials**

Supplementary materials to this article can be found in SI-PtBi2_manuscript_SciBull_short.docx

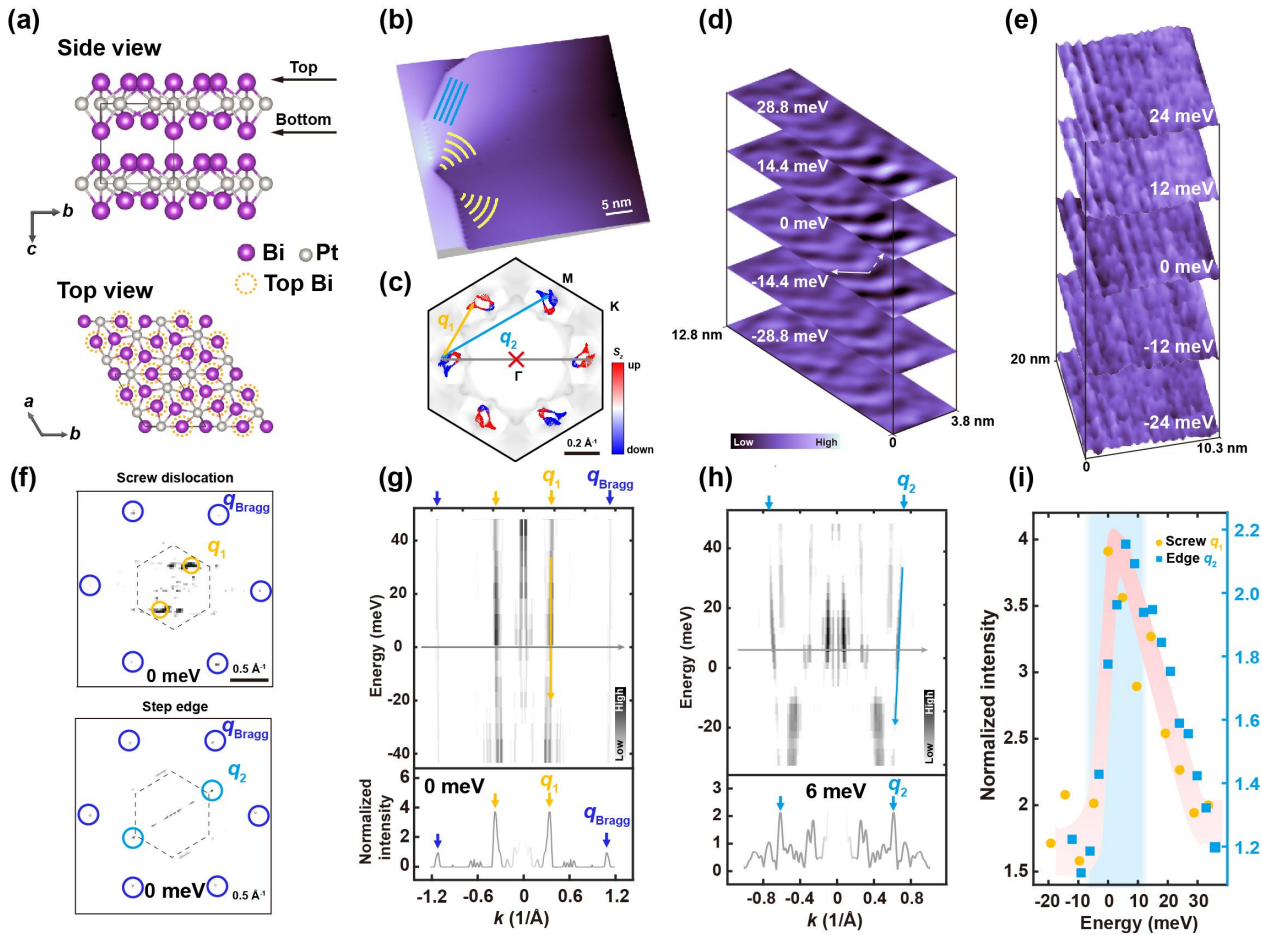

**Fig. 1.** Crystal structure of γ-PtBi$_2$ and quasiparticle interference patterns measured at 4.2 K. (a) Top panel: Side view of γ-PtBi$_2$ crystal. Bottom panel: Top view of γ-PtBi$_2$ crystal. (b) Topography of γ-PtBi$_2$ with several step-edges and screw dislocations (40 nm × 40 nm; set point, $V_s$ = -100 mV, $I_t$ = 20 pA). These edges and screw dislocations can induce quasiparticle interferences, as illustrated by sky blue and yellow ripples, respectively. (c) The comparison between the scattering wave vector obtained from the experiment and the calculated Fermi surface with spin texture. The arrows around the Fermi arcs represent the in-plane spin components, and the colors of these arrows denote the out of plane spin component. (d) and (e) d$I$/d$V$ mappings taken on a screw dislocation (3.8 nm × 12.8 nm; set point, $V_s$ = -120 mV, $I_t$ = 800 pA) and step edge (10.3 nm × 20 nm; set point, $V_s$ = -120 mV, $I_t$ = 800 pA), respectively. (f) FFT image of the d$I$/d$V$ mapping at the bias voltage of 0 mV. (g) and (h) Energy-dependent line profiles of (f) along the $q_1$ and $q_2$ direction. The intensity of each profile is normalized by the corresponding intensity of the Bragg peak. Lower panel: The line profiles at 0 meV (denoted by gray arrow) show the strong signals of $q_1$ and $q_2$. (i) Energy-dependent intensity of $q_1$ and $q_2$. The curve is taken along the yellow arrow in the upper panel of (g) and (h).

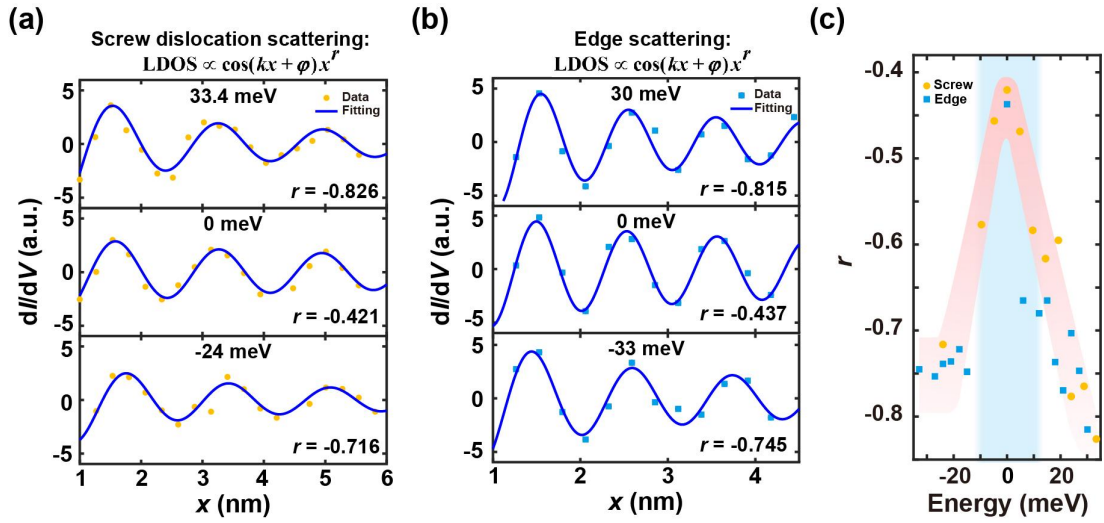

**Fig. 2.** Spatial decay of the LDOSs oscillations. (a) and (b) LDOSs oscillations at different energies originated from the screw dislocation scattering and the edge scattering, respectively. Both measurements are conducted at 4.2 K. (c) Energy dependence of the decay index of $q_1$ and $q_2$.